\documentclass[12pt]{iopart}
\usepackage{iopams,graphicx}

\begin{document}

\title{Diffusive spreading and mixing of fluid monolayers}

\author{
M N Popescu\dag\footnote[3]{To whom correspondence should be addressed}
, S Dietrich\dag~and G Oshanin\dag\ddag}

\address{\dag\ Max-Planck-Institut f{\"u}r Metallforschung,
Heisenbergstr. 3, 70569 Stuttgart, Germany, and\\
Institut f{\"u}r Theoretische und Angewandte Physik,
Universit{\"a}t Stuttgart, Pfaffenwaldring 57, 70569 Stuttgart,
Germany}

\address{\ddag\ Laboratoire de Physique Th{\'e}orique des Liquides,
Universit{\'e} Paris 6, 4 Place Jussieu, 75252 Paris, France}

\eads{\mailto{popescu@mf.mpg.de}, \mailto{dietrich@mf.mpg.de},
\mailto{oshanin@lptl.jussieu.fr}}

\begin{abstract}
The use of ultra-thin, i.e., monolayer films plays an important
role for the emerging field of nano-fluidics. Since the
dynamics of such films is governed by the interplay between
substrate-fluid and fluid-fluid interactions, the transport of
matter in nanoscale devices may be eventually efficiently
controlled by substrate engineering.
For such films, the dynamics is expected to be captured by
two-dimensional lattice-gas models with interacting particles.
Using a lattice gas model and the non-linear diffusion
equation derived from the microscopic dynamics in the continuum
limit, we study two problems of relevance in the context of
nano-fluidics. The first one is the case in which along the
spreading direction of a monolayer a mesoscopic-sized obstacle
is present, with a particular focus on the relaxation of the
fluid density profile upon encountering and passing the obstacle.
The second one is the mixing of two monolayers of different
particle species which spread side by side following the merger
of two chemical lanes, here defined as domains of high affinity
for fluid adsorption surrounded by domains of low affinity for
fluid adsorption.
\end{abstract}




\section{Introduction}
\label{intro}
There are substantial efforts to miniaturize chemical processes by
using microfluidic systems. The ``lab on a chip concept''
integrates a great variety of chemical and physical processes into a
single device in a similar way as an integrated circuit incorporates
many electronic devices into a single chip \cite{giordano01}. These
microfluidic devices do not only allow for cheap mass production
but they can operate with much smaller quantities of reactants and
reaction products than standard laboratory equipments. This is
particularly important for solutions containing rare and expensive
substances, such as certain biological materials, and for toxic or
explosive components \cite{mitchell01}.
Even though most microfluidic devices available today have micron
sized channels, further miniaturization is leading towards the
nano-scale \cite{mitchell01,stone01}. Besides meeting technical
challenges, new theoretical concepts are needed to understand the
basic physical processes underlying this new technology
\cite{dietrich98,zhao02,dietrich_jpcm_05}. Whereas the ultimate
limits for miniaturization of electronic devices are set by quantum
fluctuations, in a chemical chip these limits are determined by
thermal fluctuations and can be explored by methods of classical
statistical mechanics.

At the sub-micron thickness scale, recent experiments of liquid
spreading on atomically smooth surfaces
\cite{Heslot_prl,Leiderer_92,Voue_lang}, performed with volumes of
the order of nano-liters, have clearly shown by means of dynamic
ellipsometry or X-ray reflectivity measurements that precursor films
with {\it molecular thickness} and
{\it macroscopic extent} advance in front of the macroscopic liquid
wedge of the spreading drop. ( Thin, i.e., of the order of
100 nm, precursor films spreading ahead of the macroscopic droplet
have also been observed experimentally \cite{kavehpour03}.)
The occurrence of molecularly thin precursor films with
a similar spreading dynamics has been evidenced very recently also
for immiscible metal systems in three regimes: solid drops with a
solid film, solid drops with a liquid film, and liquid drops with a
liquid film \cite{Garoff_01,Garoff_04}. Theoretical work
(see Refs. \cite{deGennes,Oshanin_jml,Burlatsky_prl96,Abraham_02}
and references therein) combined with an impressive number of
Molecular Dynamics (MD) and Monte Carlo (MC) simulations
(see Ref.~\cite{DeConinck_96} and references therein) addressed
the mechanisms behind the extraction and the experimentally
observed $t^{1/2}$ asymptotic time dependence of the linear extent
of the precursor films on chemically homogeneous substrates
\cite{Heslot_prl,deGennes,Burlatsky_prl96,popescu_04}.
This led to a good understanding of the spreading dynamics and of the
intrinsic morphology of the films. Based on these  results, more
complicated issues can be addressed such as, e.g., the spreading
behavior of monolayers exposed to chemically patterned
substrates \cite{popescu_03,popescu_05}, or the question of mixing of
different fluids at the nano-scale that we shall present below.

The organization of the paper is as follows. In Sec.~\ref{model} we
briefly present the lattice gas model of interacting particles and
discuss the rules defining the microscopic dynamics and the nonlinear
diffusion equation derived from it in the continuum limit.
Section~\ref{obstacle} is devoted to a qualitative discussion of
the results obtained for the case of monolayer spreading in the
presence of  mesoscopic obstacles, with a particular focus on the
relaxation of the density profile upon encountering and passing
the obstacle. In Sec.~\ref{mixing} we discuss the mixing of two
species during spreading of monolayers following the merger of two
chemical lanes, and we conclude with a brief summary of the results
in Sec.~\ref{summary}.

\section{Fluid monolayers on homogeneous substrates}
\label{model}

Recently, we have studied the structure of a monolayer, which
extracts from a reservoir \cite{popescu_04} and spreads on a flat,
chemically homogeneous substrate, by using a lattice gas model of
interacting particles as proposed in
Refs. \cite{Oshanin_jml,Burlatsky_prl96}.
Since we shall use this model as a starting point for our present
study, for clarity and further reference we briefly describe the
defining rules of the model and the non-linear diffusion
equation obtained from the microscopic dynamics within the
continuum limit. A thorough analysis of this model is presented in
Ref. \cite{popescu_04}.\\
{\bf (a)} We choose a homogeneous substrate such that the spreading
occurs in the $x-y$ plane. The half-plane $x < 0$ is occupied by a
reservoir of particles at fixed chemical potential which maintains
at its contact line with the substrate --- positioned at the line
$x =0$ --- an {\it average} density $C_0$ (defined as the number
of particles per unit length in the transversal $y$ direction). At
time $t=0$, the half-plane $x>0$ is empty. There is no imposed flow
of particles from the reservoir pushing the extracting film.\\
{\bf (b)} The substrate-fluid interaction is modeled as a periodic
potential forming a lattice of potential wells with coordination
number $z$ ($z = 4$ for a square lattice) and lattice constant $a$.
The particle motion proceeds via activated jumps between
nearest-neighbor wells; evaporation from the substrate is not
allowed. The activation barrier $U_A$ determines the jumping rate
$\Omega = \nu_0 \exp[-U_A/k_B T]$, where $\nu_0$ is an attempt
frequency defining the time unit, $k_B$ is the Boltzmann constant,
and $T$ is the temperature.\\
{\bf (c)} The pair interaction between fluid particles at distance
$r$ is taken to be hard-core repulsive at short range, preventing
double occupancy of the wells, and attractive at long range,
$-U_0 /r^6$ for $r \geq 1$, resembling a Lennard-Jones type
interaction potential. Here and in the following all distances are
measured in units of the lattice constant $a$ and therefore are
dimensionless. The selection of the nearest-neighbor well
into which a particle attempts to jump, i.e., the probability
$p(\bi{r} \to \bi{r'};t)$ that a jump from location $\bi{r}$ will be
directed toward the location  $\bi{r'}$, is biased by the
fluid-fluid energy landscape and is given by
\begin{equation}
p(\bi{r} \to \bi{r'};t) =
\frac
{
\exp \bigl\{ \frac{\beta}{2}
[\tilde U(\bi{r};t)-\tilde U(\bi{r'};t)]\bigr\}
}
{
Z(\bi{r};t)
},
\label{prob}
\end{equation}
where
$Z(\bi{r};t) =
\displaystyle{
\sum_{\bi{r'}, |\bi{r'} -\bi{r}|=1}
\exp \biggl\{ \frac{\beta}{2}
[\tilde U(\bi{r};t)-\tilde U(\bi{r'};t)]\biggr\}
}
$
is the normalization constant and $1/\beta~=~k_B T$,
\begin{equation}
\tilde U(\bi{r};t) = -U_0 \sum_{\bi{r'}, 0 < |\bi{r'} -\bi{r}|
\leq 3} \frac{\eta(\bi{r'};t)} {|\bi{r} -\bi{r'}|^6},
\label{potential}
\end{equation}
and $\eta(\bi{r'};t) \in \{0,1\}$ is the occupation number of the
well at $\bi{r'}$ at the time $t$. The summation in
Eq.~\eref{potential} has been restricted to three lattice units
for computational convenience. This corresponds to the cut-off
generally used in Molecular Dynamics simulations for algebraically
decaying Lennard-Jones pair-potentials. The rates
\begin{equation}
\omega_{\bi{r} \to \bi{r'};t} = \Omega p(\bi{r} \to \bi{r'};t)
\label{rate}
\end{equation}
for the transitions from $\bi{r}$ to neighboring sites $\bi{r'}$
satisfy
\begin{equation}
\sum_{\bi{r'}, |\bi{r'} -\bi{r}| = 1} \omega_{\bi{r} \to \bi{r'};t}
\equiv \Omega.
\label{total_rate}
\end{equation}
Thus for any given particle at any location the total rate of
leaving a potential well is determined only by the fluid-solid
interaction characterized by $U_A$, is time-independent, and
equals $\Omega$.

Neglecting all spatial and temporal correlations, i.e., assuming
that averages of products of occupation numbers $\eta(\bi{r};t)$
are equal to the corresponding products of averaged occupation
numbers $\rho(\bi{r};t) = \langle \eta(\bi{r};t)\rangle$, where
$\langle \dots \rangle$ denotes the average with respect to the
corresponding probability distribution
$\mathcal{P}(\{\eta(\bi{r};t)\})$ of a configuration
$\{\eta(\bi{r};t)\}$, one can formulate a mean-field master equation
for the local occupational probability, i.e., the number density
$\rho(\bi{r};t)$ \cite{popescu_04}. In the continuum limit of space
and time ($\Delta t \to 0$, $a \to 0$, $\Omega^{-1} \to 0$,
$D_0 = \Omega a^2/4$ finite) for the master equation, by taking
Taylor expansions for $p(\bi{r} \to \bi{r'})$ and $\rho(\bi{r'};t)$
around $\bi{r}$ and keeping terms up to second-order spatial
derivatives of the density $\rho(\bi{r};t)$ \cite{popescu_04,Leung},
one obtains the following nonlinear and {\it nonlocal} equation for
$\rho(\bi{r};t)$ \cite{Giacomin,Vlachos}:
\begin{equation}
\partial_t \rho = D_0 \nabla \left[\nabla \rho +
\beta \rho \,(1-\rho) \nabla U \right]+\mathcal{O}(a^2)
\label{pde_rho}
\end{equation}
where
\begin{equation}
U(\bi{r};t) \equiv \langle \tilde U(\bi{r};t)\rangle =
-U_0 \sum_{\bi{r''},\,0 < |\bi{r''}-\bi{r}| \leq 3}
\frac{\rho(\bi{r''};t)}{|\bi{r''} -\bi{r}|^6}
\label{potential_rho}
\end{equation}
is replacing $\tilde U(\bi{r};t)$ in the definition \eref{prob} for
$p(\bi{r} \to \bi{r'})$.

Being nonlinear and, due to the term involving the interaction
potential $U(\bi{r};t)$, nonlocal, Eq.~\eref{pde_rho} cannot be
solved analytically and in most of the cases even the computation of
a numerical solution is a difficult task. However, assuming that the
density $\rho(\bi{r};t)$ is a slowly varying function of the spatial
coordinates the potential $U(\bi{r};t)$ may be expanded as
\begin{eqnarray}
U(\bi{r};t) &= -U_0 \sum_{\bi{r'},\,0 < |\bi{r'}-\bi{r}| \leq 3}
\frac{\rho(\bi{r'};t)}{|\bi{r'} -\bi{r}|^6}\nonumber \\
&\simeq
-U_0 \rho(\bi{r};t)\sum_{\bi{r'},\,0 < |\bi{r'}-\bi{r}| \leq 3}
\frac{1}{|\bi{r'} -\bi{r}|^6} + \mathcal{O}(a^2)\,,
\label{expand_pot}
\end{eqnarray}
which leads to the {\it local} equation
\begin{equation}
\partial_t \rho = D_0 \nabla
\{\left[1 - g\,W_0 \rho (1-\rho)\right] \nabla \rho \} +
{\cal O} (a^2),
\label{pde_rho_g}
\end{equation}
where $W_0 = \beta U_0$, and
$g = \displaystyle{\sum_{1 \leq |\bi{r}| \leq r_c} |\bi{r}|^{-6}}$
is a geometrical factor depending on the lattice type (e.g., square,
triangular, etc.) and on the cut-off range of the potential. For
the present case of a square lattice and a cut-off at $r_c = 3$ one
has $g \simeq 4.64$.

Rescaling time as $t \to \tau = D_0 t$ and defining an effective
diffusion coefficient
\begin{equation}
D_{e}(\rho) = 1 - g\,W_0 \rho (1-\rho),
\label{D_rho}
\end{equation}
Eq.~\eref{pde_rho_g} may be written in the usual form of a diffusion
equation:
\begin{equation}
\partial_\tau \rho =
\nabla \left[D_{e}(\rho)\nabla \rho\right]+
{\cal O} (a^2) \,.
\label{diffu}
\end{equation}
The functional form of $D_e(\rho)$ (Eq.~\eref{D_rho}) implies
that for $W_0 > 4/g$ there will be values $\rho_i$ of the density
for which $D_{e}(\rho_i) < 0$. For parameters such that $W_0 < 4/g$,
Eq.~\eref{diffu} is a proper diffusion equation (though non-linear),
while for $W_0 > 4/g$ instabilities are expected in the range of
densities for which $D_{e}(\rho_i) < 0$, i.e., for
$\rho_i\,\in\,\left(\rho_\alpha^-\,\,, \,\rho_\alpha^+ \right)$
where
\begin{equation}
\rho_\alpha^\pm =
\frac{1}{2}\left(1\pm\sqrt{1-\frac{4}{gW_0}}\,\,\right)\,.
\label{rho_alph_pm}
\end{equation}
It is known \cite{Giacomin,Elliott,Witelski_95,Witelski_96} that
these instabilities lead to discontinuities in the density profile
(``shocks''), i.e., they correspond to the formation of sharp
interfaces. For the model defined by the rules {\bf (a)-(d)}, the
value for the threshold interaction strength for which such
interfaces emerge is predicted by the continuum theory as
$W_0^{(t)} = 4/g \simeq 0.86$, which is significantly smaller than
the lower bound estimate $W_0^{(t)} > 1$ from KMC simulations.
We attribute this to the mean-field character of the derivation of
the continuum equation. Therefore it is necessary to include
particle-particle correlations into the mean-field description.
Since the dynamics is possible only by jumps into empty sites, one
can argue that for $z =4$ the summation in $g$ should include at
most three contributions from nearest neighbor sites. This leads to
$g \simeq 3.64$ and an estimate for the threshold interaction
$W_0^{(t)} \simeq 1.1$, in good agreement with the KMC results. For
the rest of the analysis we shall use this corrected value of $g$.
Additional support for this corrected value is provided by the
analysis of the density profiles \cite{popescu_04}.

\begin{figure}[htb!]
\begin{minipage}[c]{.75\textwidth}
\hspace{+1.4in}%
\includegraphics[width=.95\textwidth]{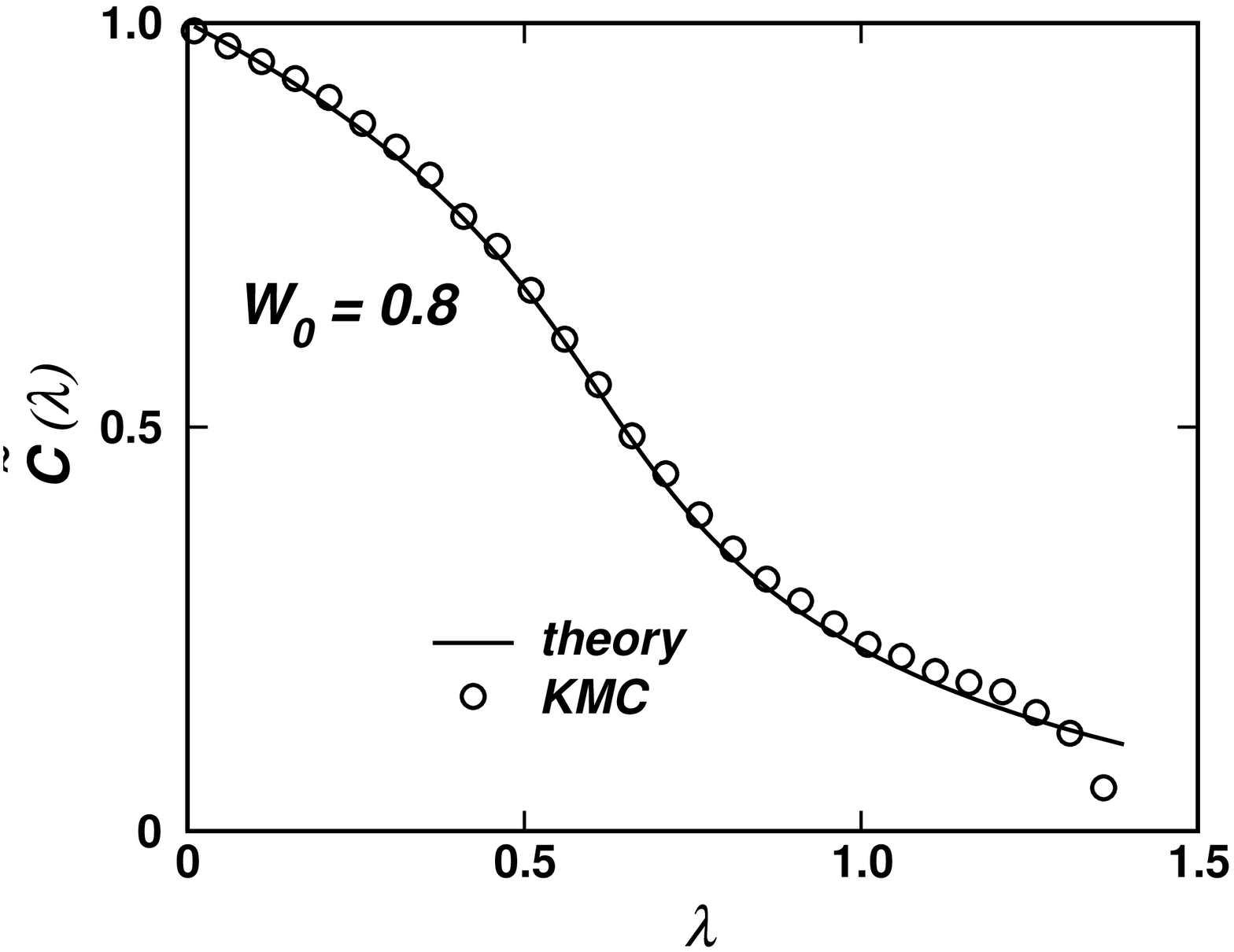}%
\raisebox{.388\textwidth}
{
\hspace{-2.16in}%
\includegraphics[width=.40\textwidth]{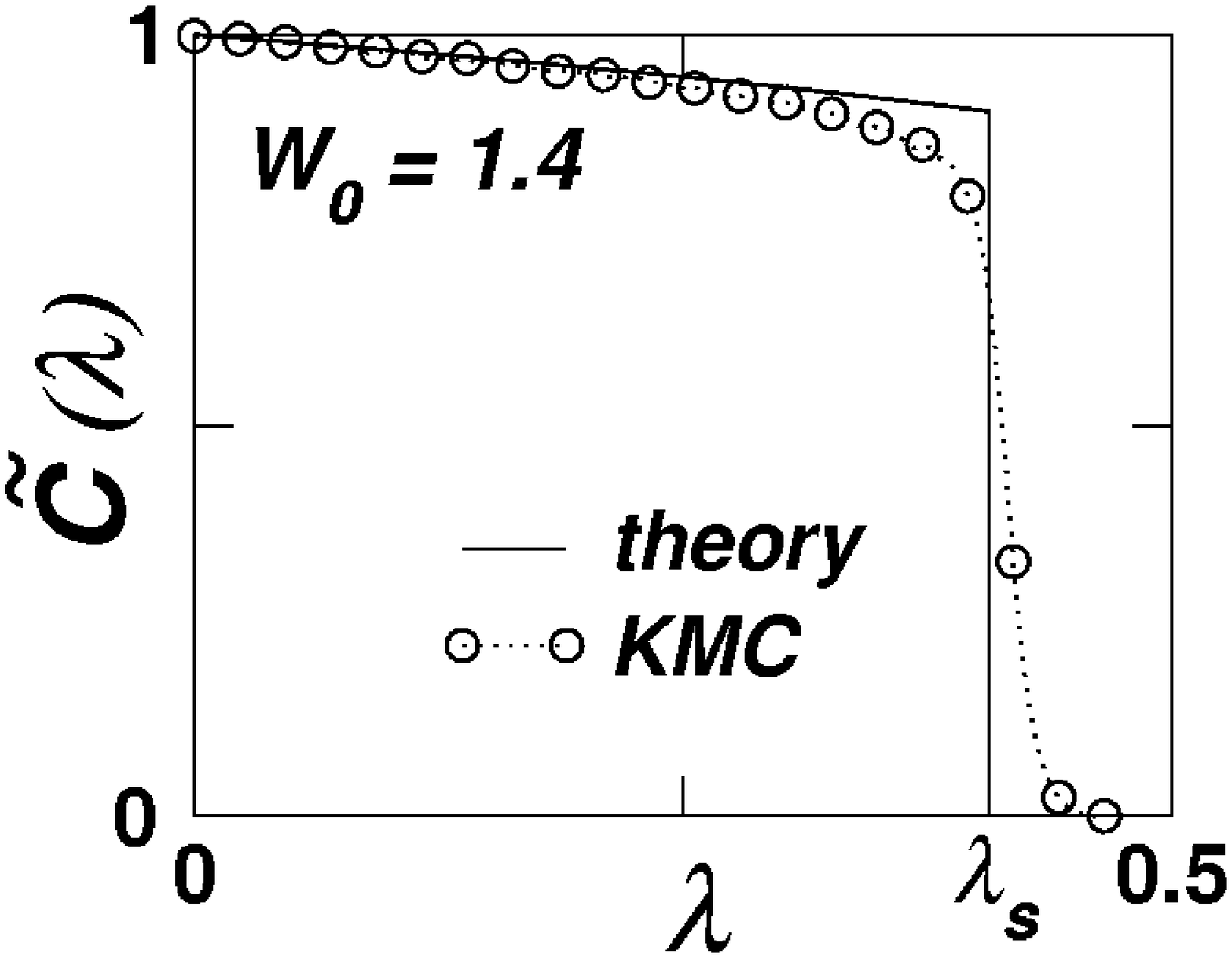}%
}%
\end{minipage}
\caption
{
\label{fig1}
Asymptotic scaling solution $\rho(x,t) = \tilde C(\lambda)$ for
$W_0 = 1.0$ (regular solution) and (see the inset) $W_0 = 1.4$
(shock solution) with $\lambda=x/\sqrt{D_0 t}$. The theoretical
results are obtained from Eq.~\eref{diffu} (solid lines). The
KMC results (open circles) correspond to the time
$t = 2 \times 10^6$ (in units of $\nu_0^{-1}$) which is close to
the asymptotic limit (see Ref.~\cite{popescu_04}). $\lambda_s$
denotes the position of the discontinuity as obtained by analytical
theory~\cite{popescu_04}.
}
\end{figure}

The constraint, that the reservoir keeps the mean density at $x=0$
at a fixed value $C_0$, implies the boundary condition
\begin{equation}
\rho(x=0,y;t) = C_0\,.
\label{BC_x0}
\end{equation}
Depending on the particular system under study, additional boundary
conditions may have to be satisfied. For example, for the case
studied in Ref.~\cite{popescu_04}, in the absence of formation of
interfaces, i.e., for interactions $W_0 < W_0^{(t)}$ and for large
times, the density on the advancing edge $X(t)$ could be considered
as fixed and equal to $C_1$,
\begin{equation}
\rho(x=X(t),y;t) = C_1\,,
\label{BC_Xt}
\end{equation}
where $C_1 = 0.11$ as inferred from the kinetic Monte Carlo (KMC)
simulations. (This boundary condition (Eq.~\eref{BC_Xt}) naturally
occurred also in the theory of Burlatsky {\it et al}
\cite{Burlatsky_prl96}.)
For that system, the absence of boundaries along the $y-$direction
and the $y$-independence of the boundary conditions at $x=0$ and
$x = X(t)$ leads to an effectively one-dimensional problem and to a
scaling solution $\rho(x,t) = \tilde C(\lambda = x/\sqrt{t})$.
The analysis of Eq.~\eref{diffu} depends on whether
$W_0 < W_0^{(t)}$ or $W_0 > W_0^{(t)}$. As shown in
Ref.~\cite{popescu_04}, in both cases the solutions are in
excellent agreement with those obtained from KMC simulations; typical
results are shown in Fig.~\ref{fig1}.
While for the liquid-on-solid systems mentioned in the Introduction
these intrinsic density profiles have not been measured yet,
data of such density profiles are available for the immiscible
metal systems studied in Ref.~\cite{Garoff_01,Garoff_04} and they
are in at least good qualitative agreement with the theoretical ones.
Since the present model appears to provide a simple but realistic
description of a fluid monolayer spreading on a homogeneous
substrate, it is natural to use it as a starting point to
address more complex problems, such as the spreading of monolayers
on \textit{designed} chemically heterogeneous substrates, or the
mixing of monolayers.

\section{Diffusive spreading around mesoscopic obstacles}
\label{obstacle}

\begin{figure}[htb!]
\begin{minipage}[c]{.3\textwidth}
\includegraphics[width = 0.9\textwidth]{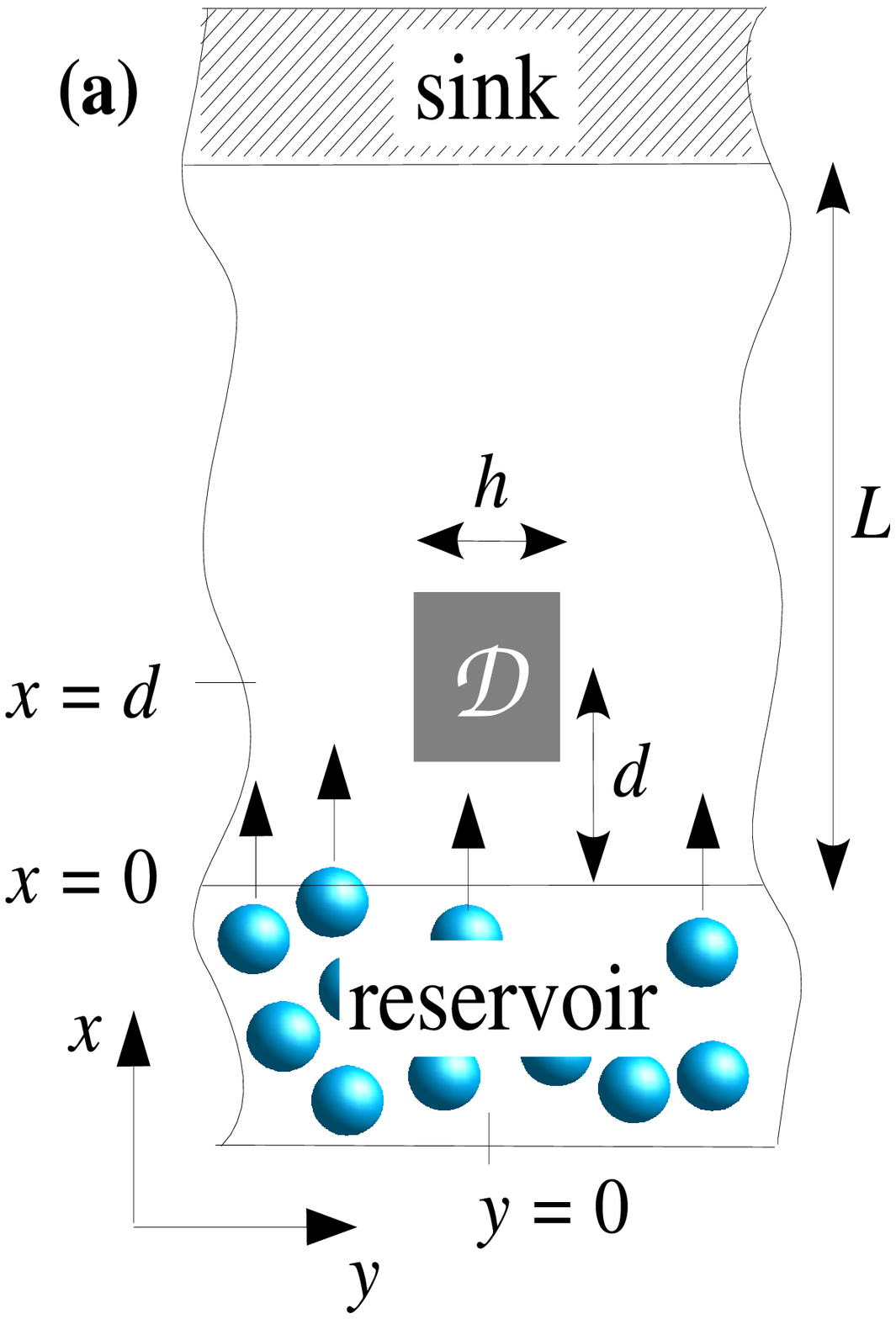}
\end{minipage}%
\begin{minipage}[c]{.3\textwidth}
\centerline{(b)}
\includegraphics[width = .99\textwidth]{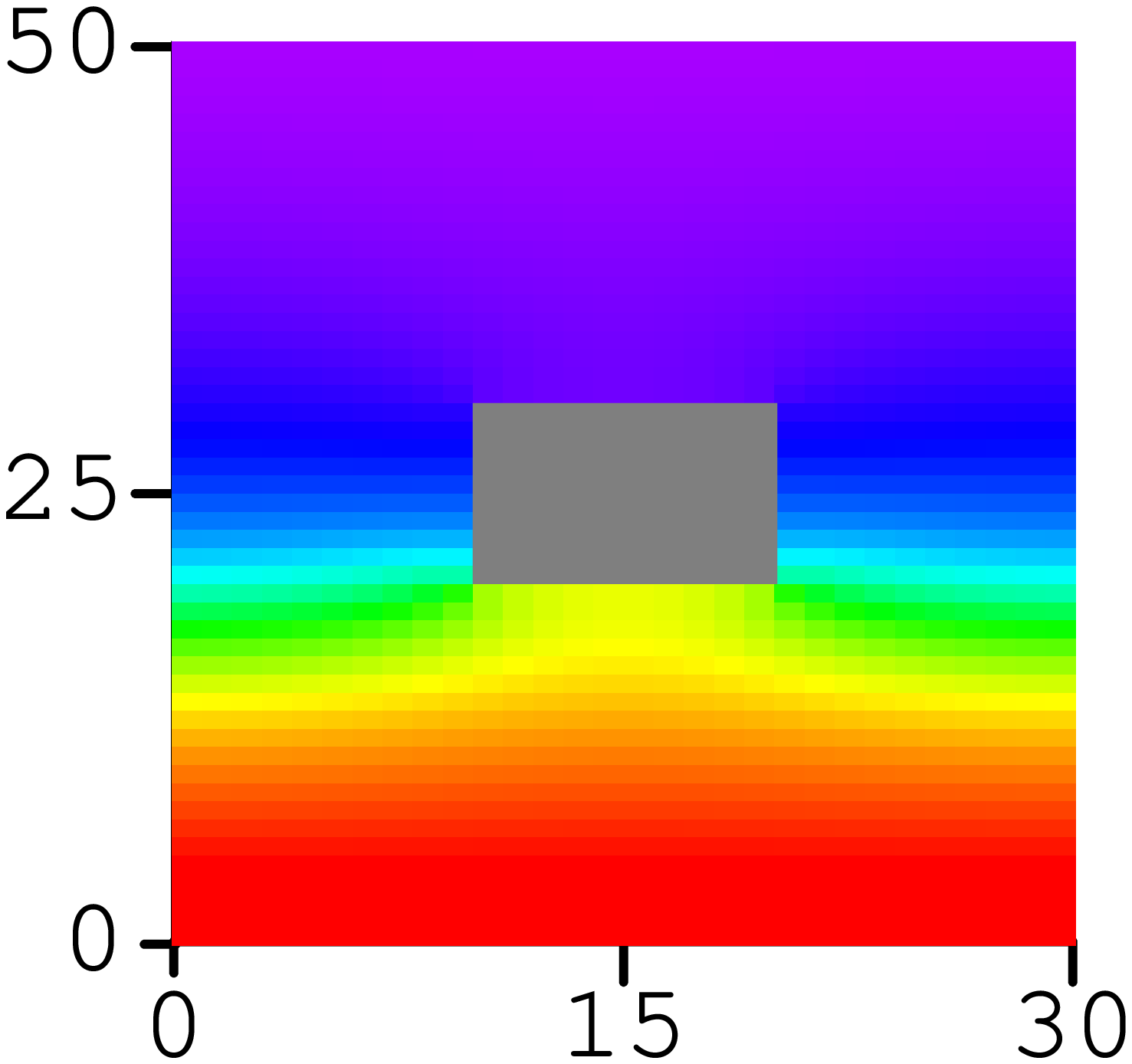}
\end{minipage}%
\begin{minipage}[c]{.3\textwidth}
\centerline{(c)}
\includegraphics[width = .99\textwidth]{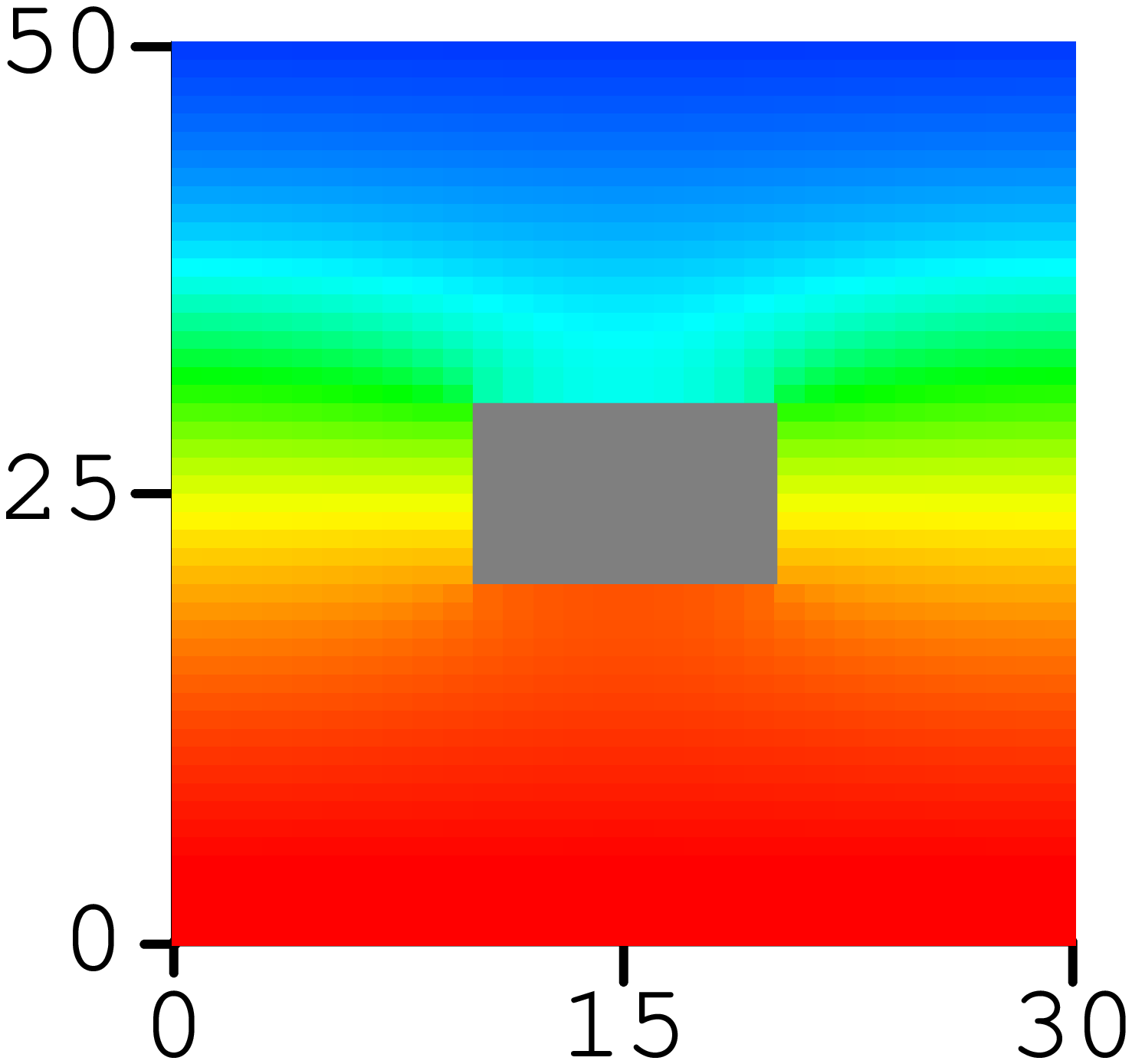}
\end{minipage}%
\begin{minipage}[c]{.1\textwidth}
$\rho = $\hfill\\
\includegraphics[width = 0.8\textwidth,height=1.4in]{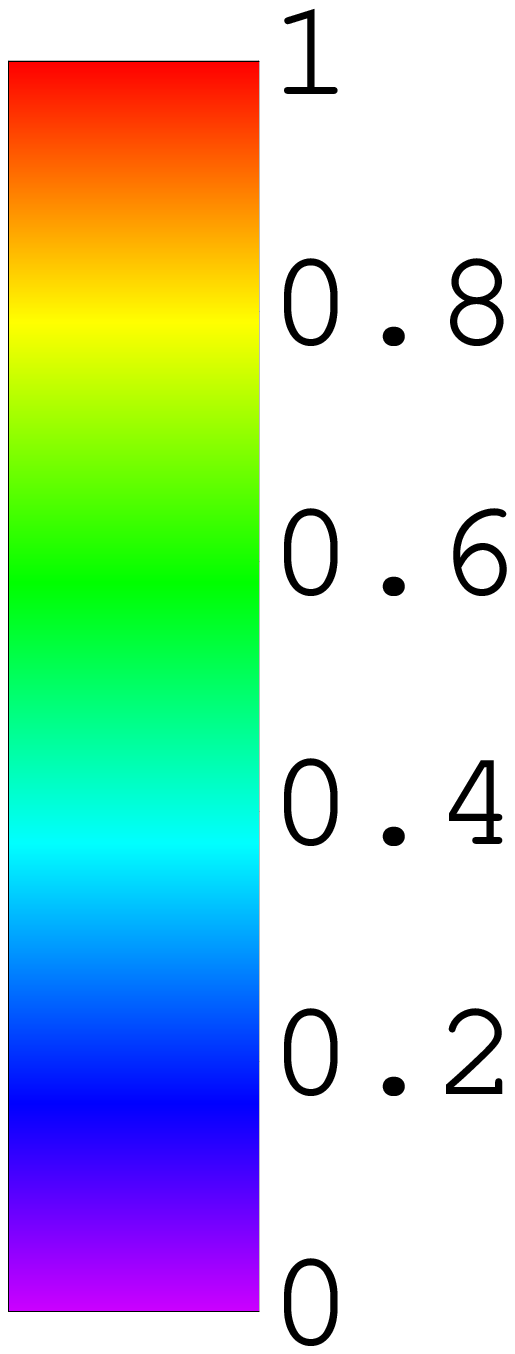}
\end{minipage}
\caption
{
\label{fig2}
(a) Schematic drawing of a substrate with high affinity for fluid
adsorption patterned with a square domain $\cal D$
[length $h$, centered at $(x = d \geq h/2, y = 0)$] of very low
affinity for fluid adsorption obstructing the spreading. The
substrate is in contact with a reservoir of particles located at
$x \leq 0$ from which a monolayer is extracted and spreads along
the $x$ direction. A sink of particles is located at $x = L$. There
are periodic boundary conditions in the $y-$direction.
(b),(c) Density profiles in the vicinity of the obstacle from
numerical integration of Eq.~\eref{diffu} with initial and boundary
conditions given by Eqs.~\eref{IC}-\eref{BC_obstacle}, respectively,
for a spreading monolayer whose edge just encounters an obstacle
($\tau = 200$) (b) and has just passed the obstacle ($\tau = 2000$)(c).
The parameters used in the simulations are $C_0 = 1.0$, $W_0 = 0.9$,
$d = 20$, $h = 10$, $L_x = 200$, $L_y = 30$. The right box shows the
color coding for the density.
}
\end{figure}

In order to apply the model described in Sec.~\ref{model} to study
the spreading of a monolayer around a mesoscopic-size obstacle, we
add to the rules {\bf (a)-(d)} in Sec. II the following ones:\hfill\\
{\bf (e)} The obstacle is taken to be a square-shaped domain
$\cal D$ of side length $h$ centered at $(x = d \geq h/2, y = 0)$
[see also Fig.~\ref{fig2}(a)]. This domain is composed of sites
with very low affinity for the fluid particles. The activation
barrier $U_{\cal D}$ for jumps from sites outside $\cal D$ to those
inside $\cal D$ is taken to be much larger than $U_A$, such that the
boundary ${\cal \partial D}$ of $\cal D$ acts effectively as a hard
wall.\hfill\\
{\bf (f)} A sink reservoir occupies the region $x \geq L$,
where $L \gg 1$ and $L \gg d+h$, and maintains at its contact line
with the substrate, positioned at the line $x = L$, an {\it average}
density (number of particles per unit length in the transversal $y$
direction) $C_1  = 0$.

Under these assumptions, the density profile $\rho(x,y,t)$ as the
solution of Eq.~(\ref{diffu}) fulfills the initial condition
\begin{equation}
\rho(x,y,0) = C_0 \Theta(-x),
\label{IC}
\end{equation}
where $\Theta(x)$ denotes the Heaviside step function, and the
boundary conditions
\begin{eqnarray}
\rho(0,y,t) &=& C_0, \nonumber\\
\rho(L,y,t) &=&  0, \\
\bi{j}_n \left|_{\partial {\cal D}}\right. = 0\,. \nonumber
\label{BC_obstacle}
\end{eqnarray}
The current $\bi{j}$ is given by [see Eq.~(\ref{diffu})]
\begin{equation}
\bi{j} = - D_{e}(\rho)\nabla \rho\,.
\label{current}
\end{equation}

Note that in a numerical study the system necessarily has a finite
size $L_y$ along the $y$ direction. We will use periodic boundary
conditions and sizes $L_y \gg1$ such that $h/L_y < 1$ but not
negligible (mesoscopic-size obstacle) and $L_y - h \gg r_c$, such
that the boundary $\partial {\cal D}$ of the obstacle is sufficiently
far away from the edge of the simulation box  to avoid finite-size
effects.

In Figs.~\ref{fig2}(b) and (c) we present typical results for the
density profiles in the vicinity of the obstacle obtained from the
numerical integration of Eq.~\eref{diffu} with initial and boundary
conditions given by Eqs.~\eref{IC}-\eref{BC_obstacle}, respectively,
for a spreading monolayer whose edge just encounters an obstacle
($\tau = 200$)(b) and has just passed the obstacle ($\tau = 2000$)(c).
Several conclusions can be drawn from visually inspecting
Figs.~\ref{fig2}(b) and (c).
Upon approaching the obstacle, the boundary condition of zero normal
current at the boundary of the obstacle (reflecting wall) leads
to an increase in the density in a region at the front of the
spreading monolayer, as shown by the forward bending of the
iso-density lines [see, e.g., the yellow band in Fig.~\ref{fig2}(b)].
Upon passing the front-edge of the obstacle, the iso-density lines
become straight, and once they reach the end of the obstructed region
they bend again, this time backwards. This indicates that upon passing
the obstacle the spreading tends to proceed faster in the regions far
from the obstacle, while at the obstacle the iso-density lines are
pinned until they cover the whole edge on the back of the obstacle
[see, e.g., the light blue band neighboring the green region in
Fig.~\ref{fig2}(c)]. Once this is realized, the iso-density line
detaches from the back-edge of the obstacle, and the bending slowly
relaxes  [see, e.g., the boundary between the light and dark blue
regions in the top region of Fig.~\ref{fig2}(c)], while the spreading
continues; far away from the obstacle, the iso-density lines become
again straight.

We end this section by noting that the maximum linear extent of the
region where the iso-density lines are deformed, as well as the
survival time of these deformations can be used as quantitative
measures to describe the relaxation of the density perturbations
induced by the obstacle as a function of the inter-particle
attractive interactions, as well as of the scaled size $h/L_y$ of
the mesoscopic obstacle (assuming that this is the most relevant
geometrical parameter). The results of this analysis will be the
subject of a forthcoming paper.

\section{Diffusive mixing of two fluid monolayers composed of
different species}
\label{mixing}
The model described in Sec.~\ref{model} can be used to study the
mixing of two spreading fluid monolayers composed of different
species $A$ and $B$, if one assumes that the two species interact
with the substrate in such a way that the same lattice structure of
potential wells, eventually with different depths (i.e., different
escape rates $\Omega_i$), can accommodate both types of species.
Assuming a square lattice of lattice constant $a$ and assuming the
on-site hard core repulsion between any two particles such that
double occupancy remains forbidden, the equations satisfied by the
densities $\rho_{j}(x,t)$, where $j \in \{A,\,B\}$, are obtained
from Eq.~\eref{pde_rho} by replacing $U$ with $U_{ii} + U_{ij}$,
where $U_{ii}$ is the potential due to same-species interactions
while $U_{ij}$, $j \neq i$, is the potential due to interactions
between different species, $D_0$ with $D_i=\Omega_i a^2/4$, and
changing the single-occupancy term from
$1-\rho(\bi{r},t)$ to $1-\rho_A(\bi{r},t) -\rho_B(\bi{r},t)$:
\begin{equation}
\fl \hspace{.5in}
\partial_t \rho_i = D_i \nabla \left[\nabla \rho_i +
\beta \rho_i \,(1-\rho_i-\rho_j) \nabla (U_{ii} + U_{ij}\right], ~
i,\,j \in \{A,\,B\}, ~ j \neq i\,.
\label{pde_rho_mix}
\end{equation}
We note here that there is no summation over the same indices,
and we also note that in the case of identical species, i.e.,
$D_A = D_B$ and $U_{AA} = U_{BB} = U_{AB}$, by adding
Eq.~\eref{pde_rho_mix} for $i = A$ and the corresponding one
for $i = B$ one finds that as expected the density
$\rho = \rho_A + \rho_B$ satisfies Eq.~\eref{pde_rho}.
Assuming that the long-ranged parts of the pair-interactions $A-A$,
$B-B$, and $A-B$ are of the same Lennard-Jones form (see
Sec.~\ref{model}) only with different strengths
$U_0^{(ij)}$,  $i,\,j \in \{A,\,B\}$, one can repeat the same
argument as that following Eq.~\eref{pde_rho} to reduce
Eq.~\eref{pde_rho_mix} to a \textit{local} one,
\begin{equation}
\fl \hspace{.1in}
\partial_t \rho_i = D_i \nabla \left[\nabla \rho_i -
g\, \rho_i \,(1-\rho_i-\rho_j) ( W_{ii} \nabla \rho_{i} +
W_{ij} \nabla \rho_{j} \right],
~ i,\,j \in \{A,\,B\}\,, ~ j \neq i\,,
\label{local_pde_rho_mix}
\end{equation}
where the notation $W_{ij} = \beta U_0^{(ij)}$ has been introduced.

\begin{figure}[htb!]
\begin{minipage}[c]{.3\textwidth}
\centerline{(a)\hspace*{.2in}}
\vspace*{.1in}
\includegraphics[width = 0.9\textwidth]{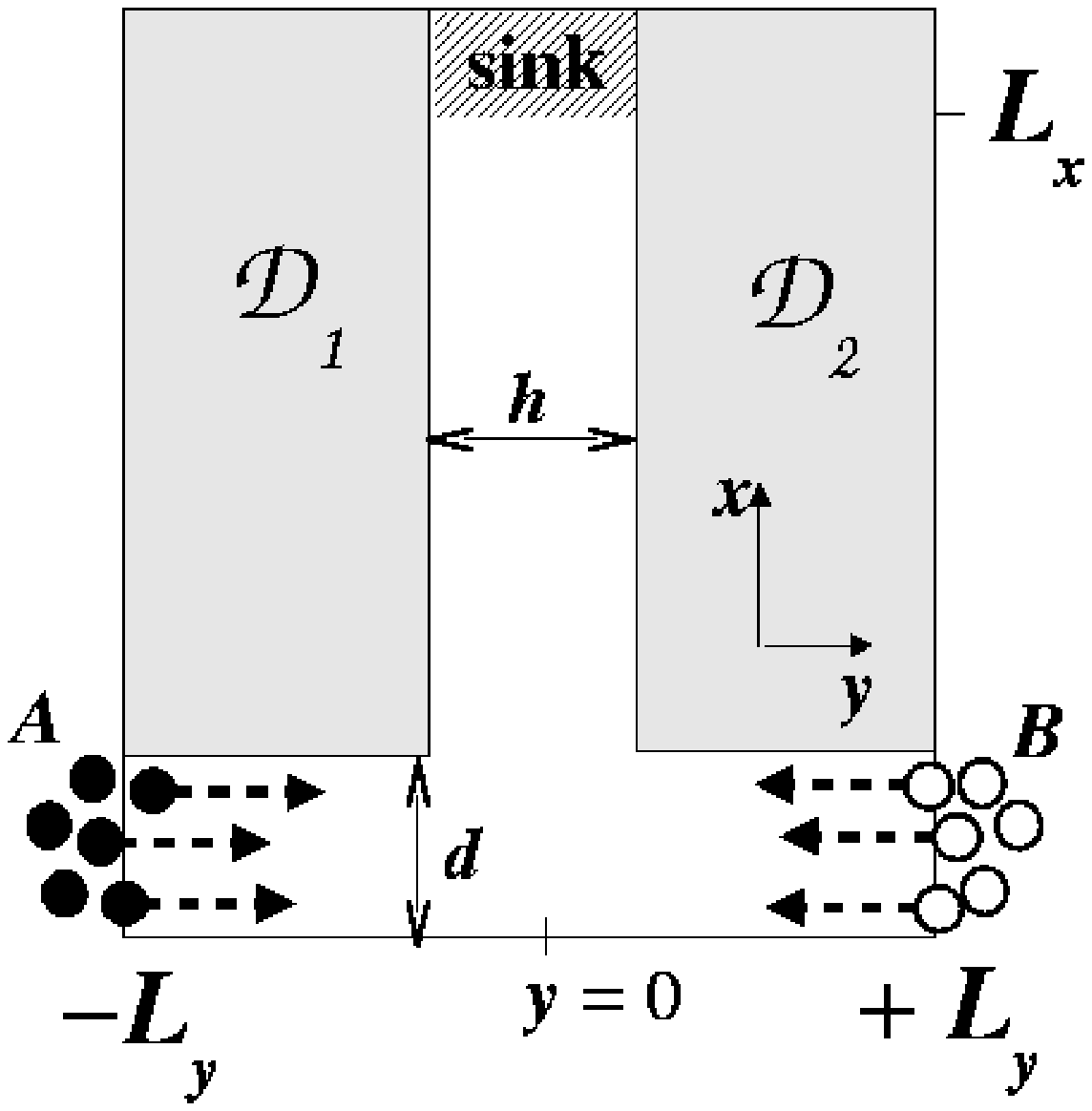}
\vspace*{1.in}
\end{minipage}%
\begin{minipage}[c]{.3\textwidth}
\centerline{\hspace*{.1in}(b)}
\includegraphics[width = 0.95\textwidth]{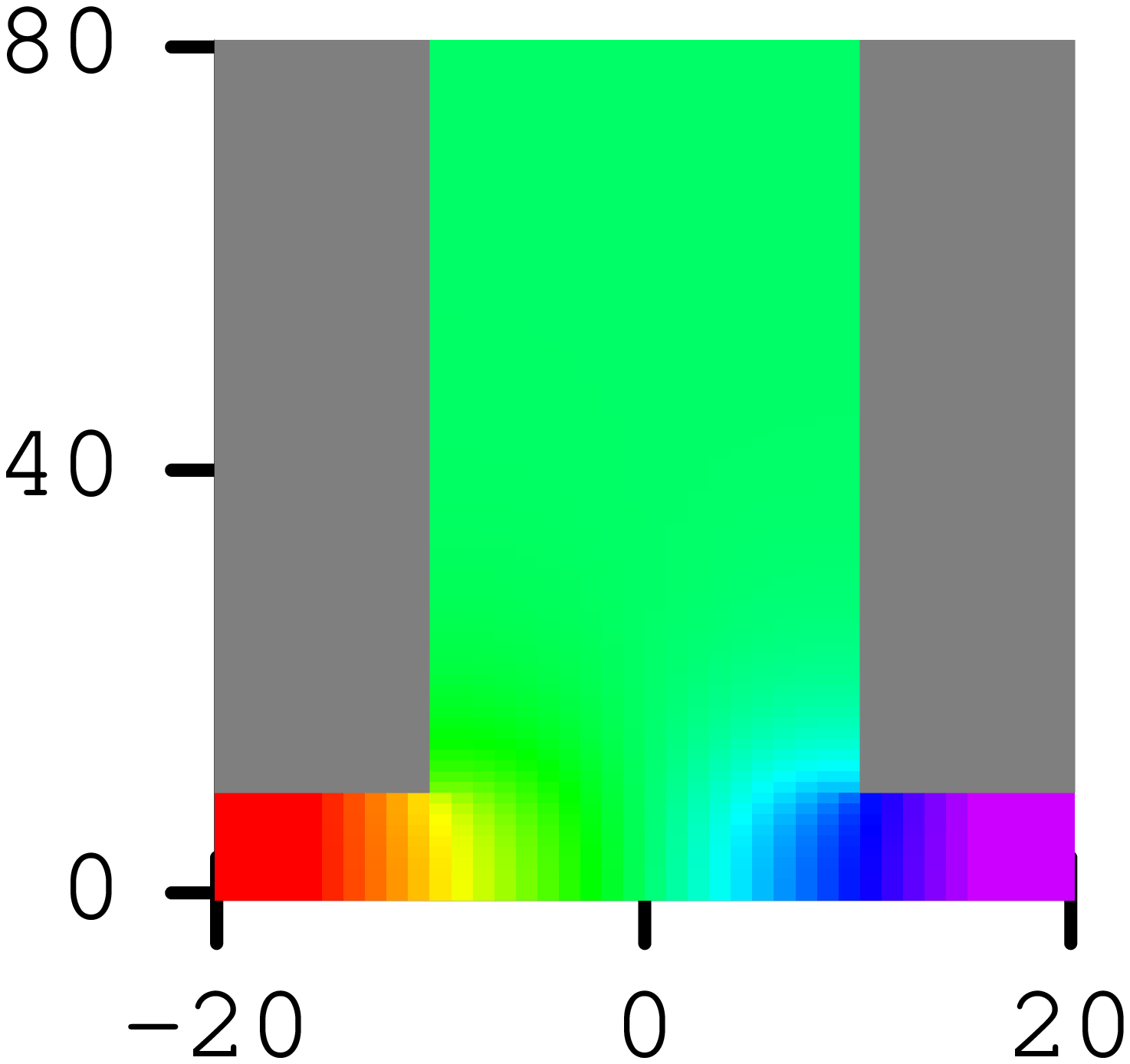}
\centerline{\hspace*{.01in}(d)}
\hspace*{-.2in}
\includegraphics[width = 1.1\textwidth]{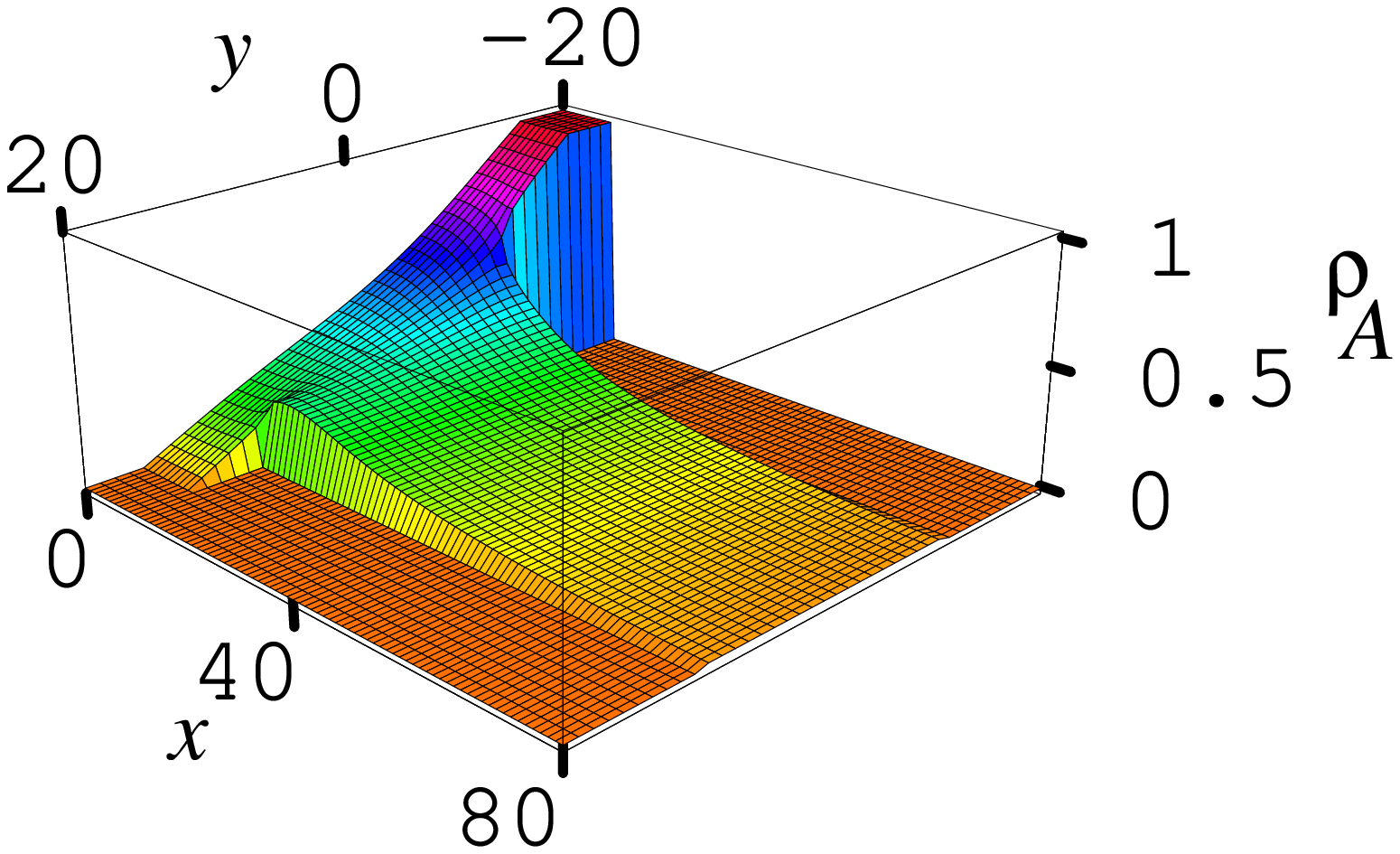}\hspace*{.7in}
\end{minipage}%
\begin{minipage}[c]{.3\textwidth}
\centerline{\hspace*{.1in}(c)}
\includegraphics[width = 0.95\textwidth]{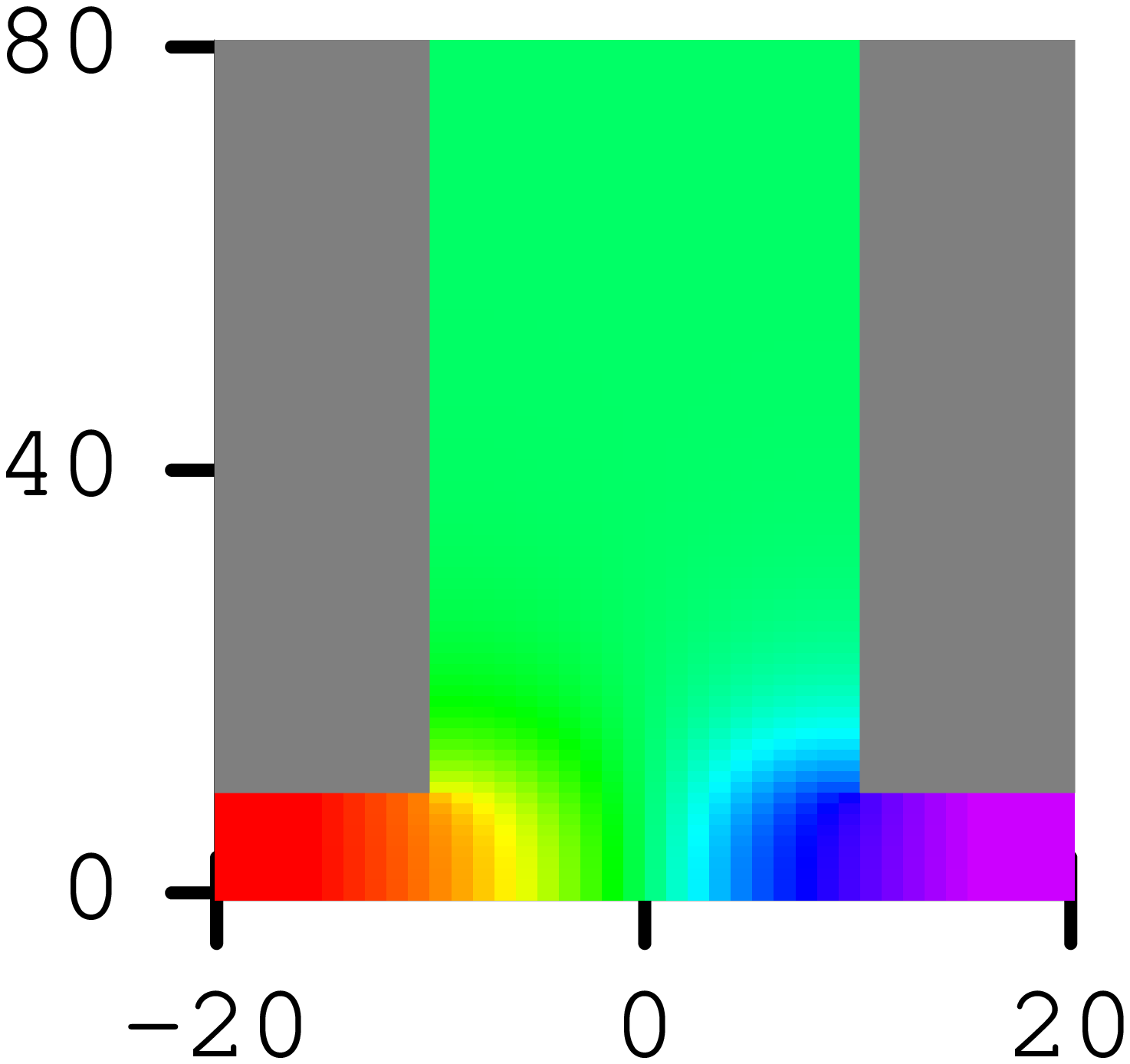}
\centerline{\hspace*{.8in}(e)}
\hspace*{.2in}
\includegraphics[width = 1.1\textwidth]{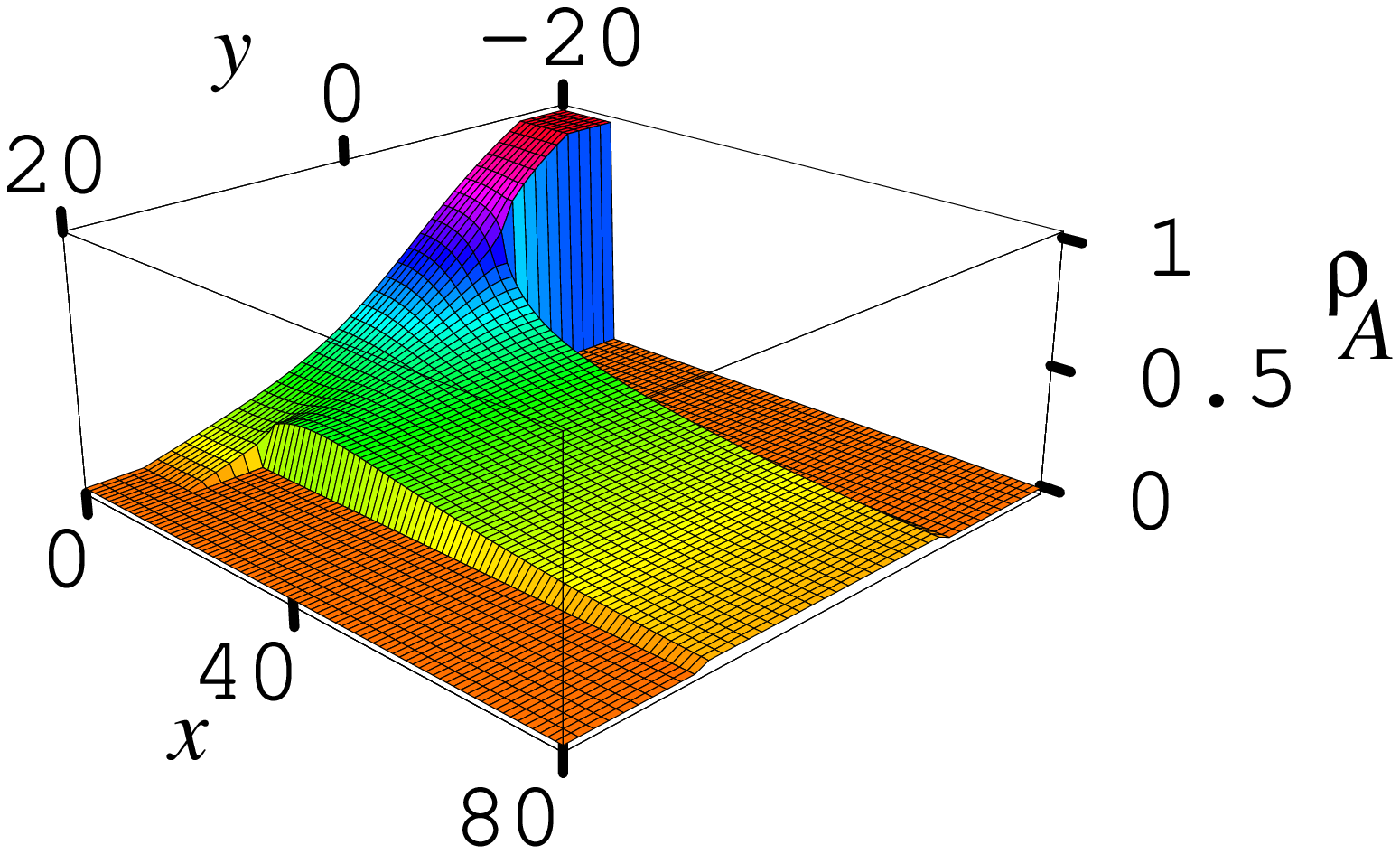}\hspace*{.7in}
\end{minipage}%
\begin{minipage}[c]{.1\textwidth}
 $c_A = $\hfill\\
 \vspace*{.1in}
\includegraphics[width = 0.8\textwidth,height=1.4in]{color_code.eps}
\vspace*{1.5in}
\end{minipage}
\caption
{
\label{fig3}
(a) Schematic drawing of a T-shaped pattern with high affinity for
a fluid in contact with two reservoirs for particles of species $A$
and $B$, located at $y = -L_y$ and $y = L_y$, respectively. The two
monolayers spread initially in opposite directions along the $y$
axis, and upon meeting they start to mix and to invade the empty
stripe of width $h$ located at $x > d, |y| \leq h/2$. The
rectangular domains ${\cal D}_1$ and ${\cal D}_2$ consist of sites
with very low affinity for fluid adsorption. A sink of particles is
located at $x = L_x$. (b), (c) Results for the mixing parameter
$c_A(\bi{r},\tau)$ near the T-junction from numerical integration of
Eq.~\eref{local_pde_rho_mix} with initial and boundary conditions
given by Eqs.~\eref{IC_mix}-\eref{BC_mix}
for the cases of attractive inter-species interaction
$W_{AB} = 0.6$ (b) and repulsive inter-species interaction
$W_{AB} = - 0.6$ (c), respectively, at $\tau = 10^3$, for
$W_{AA} = W_{BB} = 0.7$, and $C_0^A = C_0^B = 1.0$.
(d), (e) Results for the density $\rho_A(\bi{r},\tau)$ of $A$
particles for attractive inter-species interaction
$W_{AB} = 0.6$ (d) and repulsive inter-species interaction
$W_{AB} = - 0.6$ (e). The other parameters are the same as in
(b) and (c).
Note that there is no color-coding for $\rho_A(\bi{r},\tau)$.
}
\end{figure}

We consider a T-junction patterned onto a planar, rectangular
substrate of size $L_x \times 2 L_y$ as described below [see also
Fig.~\ref{fig3}(a)].
The spreading monolayers are extracted from reservoirs of
particles $A$ and $B$, respectively, which maintain constant line
densities $C_0^{A,B}$ at the lines
${\cal C}_1 = (y = -L_y, 0 \leq x \leq d)$ and ${\cal C}_2 =
(y = L_y, 0 \leq x \leq d)$, respectively. The domains
${\cal D}_1 = \{ (x,y)| y < - h/2 \wedge x > d \}$ and
${\cal D}_2 = \{ (x,y)| y > h/2 \wedge x > d \}$ represent sites
with very low affinity for the fluid particles of either type, such
that similarly to the situation in Sec.~\ref{obstacle} the
boundaries of these domains effectively act as hard walls, confining
the spreading onto the two lanes forming the inverted T-junction.
Finally, we assume that at the foot $x = L_x, |y| \leq h/2$ of the
T-junction there is a sink for particles of both species. Under
these assumptions, the functions $\rho_{A,B}(x,y,t)$ as
solutions of Eq.~\eref{local_pde_rho_mix} fulfill the initial
condition
\begin{eqnarray}
\rho_A(x,y,0) = C_0^{A}\,,\, && (x,y) \in {\cal C}_1\,,\nonumber\\
\rho_B(x,y,0) = C_0^{B}\,,\, && (x,y) \in {\cal C}_2\,,\\
\rho_{A,B}(x,y,0) = 0, &&\mathrm{otherwise}\nonumber,
\label{IC_mix}
\end{eqnarray}
and the boundary conditions
\begin{eqnarray}
\rho_A(x,y,t)\left|_{{\cal C}_1} \right. = C_0^{A},\nonumber\\
\rho_B(x,y,t)\left|_{{\cal C}_2} \right. = C_0^{B},\\
\rho_A(L_x,y,t) = \rho_B(L_x,y,t) = 0, \nonumber\\
\bi{j}^{A,B}_n \left|_{\partial {\cal D}_{1,2}}\right.
= 0\,, \nonumber
\label{BC_mix}
\end{eqnarray}
where the current $\bi{j}^A$ ($\bi{j}^B$ is obtained by exchanging
the labels $A \leftrightarrow B$) is now given by
[see Eq.~\eref{local_pde_rho_mix}]
\begin{equation}
\bi{j}^{A} = \nabla \rho_A -
g\, \rho_A \,(1-\rho_A-\rho_B) ( W_{AA} \nabla \rho_{A} +
W_{AB} \nabla \rho_{B})\,.
\label{current_mix}
\end{equation}

In the following we focus on the effect of the A-B interaction
on the dynamics of mixing of otherwise identical monolayers, i.e.,
we choose $W_{AA} = W_{BB}$, $D_A = D_B$, and $C_0^A =C_0^B$; the
results discussed in the following correspond to the particular
choice $C_0^A = C_0^B = 1$, while the parameter $D_A$ is absorbed
into the variable $\tau = D_A t$. The geometrical parameters are
fixed to $L_x  = 500$, $L_y = 20$, and $d = h = 20$. The mixing
will be characterized by the ratio
\begin{equation}
c_A(\bi{r},\tau) =
\frac{\rho_A(\bi{r},\tau)}{\rho_A(\bi{r},\tau)+\rho_B(\bi{r},\tau)}
\label{mix_param}
\end{equation}
(with the convention $c_A = 0$ if $\rho_A + \rho_B = 0$), which is
close to 1 in A-rich regions, close to zero in B-rich regions, and
close to 1/2 in regions where mixing is accomplished (i.e.,
$\rho_A \simeq \rho_B \neq 0$).

In Figs.~\ref{fig3}(b) and (c) we present typical results for
$c_A(\bi{r},\tau)$ from numerical integration of the coupled set
of equations given by Eq.~\eref{local_pde_rho_mix} with the initial
and boundary conditions Eqs.~\eref{IC_mix}-\eref{BC_mix} for the
case of attractive interactions $W_{AA} = W_{BB} = 0.7$ and
attractive inter-species interaction $W_{AB} = 0.6$ (b), respectively
repulsive inter-species interaction $W_{AB} = - 0.6$ (c). From these
figures it is clear that within the stripe forming the leg of the
T-junction there is almost perfect mixing.

Surprisingly, at first glance the result seems to be almost
independent of the sign of the A-B interaction, and there is only a
weak dependence on the strength $W_{AB}$ of this interaction, except
for the extension of the A-rich and B-rich regions near the corners of
the T-junction. However, this behavior can be easily rationalized in
view of the fact that, as shown in Ref.~\cite{popescu_04}, the structure
and dynamics of the spreading of a one-component monolayer of particles
with inter-particle attraction $W_{AA} = 0.7$ is well described also
by the "effective boundary force" theory of Burlatsky \textit{et al}
\cite{Burlatsky_prl96}, which disregards interactions within the bulk
of the monolayer. The reason for this is that the density
in the spreading monolayer is relatively low, except for the region
near the reservoir, and thus the system is too dilute to be influenced
by the inter-particle attraction. Therefore, in the initial stages of
spreading and mixing, the stripe is invaded by low density phases of
$A$ and $B$ which mix independently of their mutual interaction. This
scenario is well supported by the comparative analysis of the
corresponding density profiles $\rho_{A}(\bi{r},\tau)$ in the cases
$W_{AB} = 0.6$ and $W_{AB} = - 0.6$, respectively, shown in
Figs.~\ref{fig3}(d) and (e): almost everywhere in the stripe the
density of $A$ particles is low, in the range of 0 to 0.25, and thus
the mixture behaves as a dilute, non-interacting two-dimensional gas.
Finally, we note a weak dependence of the extension of the A-rich and
B-rich regions near the corners of the T-junction on the sign of the
$W_{AB}$ interaction [compare Figs.~\ref{fig3} (b) and (c),
respectively (d) and (e)].

The above scenario holds for all  values $W_{AA} = W_{BB}
\leq 0.9$ and $|W_{AB}| \leq W_{AA}$ that we have tested, and thus
the (tentative) conclusion is that the symmetric T-geometry would
ensure practically perfect mixing (but without any
possibility to control the spatial extent or the spatial distribution
of mixing) for two-component monolayers with a similarity of the
interactions between like species. However, further calculations should
be carried out before definite conclusions can be drawn concering this
issue. For example, for the case $W_{AA},W_{BB} > 1.1$ we expect sharp
interfaces to emerge in each of the two spreading monolayers, which
might eliminate the ''mixing through the low density front'' mechanism
discussed above. Simulations for these ranges of values for the
inter-particle interactions turned out to be extremely time consuming,
and work is still in progress to elucidate this point.

\section{Summary}
\label{summary}

A lattice gas model of interacting particles and the corresponding
nonlinear diffusion equation derived from its microscopic dynamics
in the continuum limit provide a simple but realistic description
of fluid monolayer spreading on a homogeneous substrate. Based on
previous results for spreading on a homogeneous substrate, here we
have extended this model to address two more complex problems: the
spreading of monolayers around obstacles (Sec.~\ref{obstacle}) and
the mixing of monolayers (Sec.~\ref{mixing}). These are simple
examples of spreading on chemically \textit{designed} substrates.

For the case of monolayer spreading in the presence of a mesoscopic
obstacle, the results obtained from the numerical integration of the
nonlinear diffusion equation (Eq.~\eref{diffu}) with initial and
boundary conditions given by Eqs.~\eref{IC}-\eref{BC_obstacle} show
that the iso-density lines are bent and pinned by the obstacle
during the spreading of the monolayer around it. For a fixed
geometry, a fixed density of the reservoir, and fixed
substrate-fluid and inter-particle interactions the spatial and
temporal extent of this bending in front of and behind the obstacle
can be used as measures for the relaxation of the density profile
upon passing around the obstacle.

As an example for the mixing of two species in the course of
spreading of two monolayers at the merger of two chemical lanes,
we have discussed the case of a T-junction geometry. We have
focused on the effect of the A-B inter-species interaction on the
dynamics of mixing of otherwise identical monolayers. Surprisingly,
so far our results lead to the conclusion that the symmetric
T-geometry together with the similarity of the same-species
interaction ensures practically perfect mixing
(but without any possibility to control the spatial extent or the
spatial distribution of mixing) for two monolayers of
different species, independently of the sign or the strength of the
A-B interaction. Only the extension  of the A-rich and B-rich regions
near the corners of the T-junction exhibits differences. This behavior
reflects the fact that in the initial stages of spreading and
mixing the stripe is invaded by low density phases of $A$ and $B$,
which mix independently of the inter-species interaction; the
resulting mixture is a dilute, quasi non-interacting two-dimensional
gas. Because this complete mixing has occurred at early stages, when
repulsion does not play a role, and since the continuum equation does
not contain any noise terms, demixing or segregation is not
observed in the present calculations, although it is expected to occur
for strongly repulsive A-B interactions.

\end{document}